\documentclass[aps,prl,twocolumn,reprint,
			   groupedaddress,superscriptaddress,
			   amsfonts,amssymb,amsmath,
			   citeautoscript,
			   a4paper
			   ]{revtex4-1}

\usepackage{microtype} 

\usepackage{txfonts}  
\usepackage{txfontsb} 
\usepackage{bm} 

\usepackage{xcolor}
\usepackage{graphicx}
\graphicspath{{./Figure_Plot_Matlab/}}

\usepackage{hyperref}
\hypersetup{
    colorlinks,
    linkcolor={blue!75!black!80!yellow},
    citecolor={blue!75!black!80!yellow},
    urlcolor={blue!75!black!80!yellow}
}

\usepackage[centering,hmargin=1.85cm,tmargin=31mm,bmargin=37mm]{geometry}

\color{black!88!white}

\newcommand{\sbp}{\scriptscriptstyle ||}

\makeatletter
\renewcommand{\fnum@figure}{\figurename~\thefigure\ (color online)}
\makeatother

\begin{document}
\title{\color{black!88!white} Projected-Dipole Model for Quantum Plasmonics }

\author{Wei~Yan}
\affiliation{Department of Photonics Engineering, Technical University of Denmark, DK-2800 Kgs. Lyngby, Denmark}
\affiliation{Center for Nanostructured Graphene, Technical University of Denmark, DK-2800 Kgs. Lyngby, Denmark}

\author{Martijn Wubs}
\affiliation{Department of Photonics Engineering, Technical University of Denmark, DK-2800 Kgs. Lyngby, Denmark}
\affiliation{Center for Nanostructured Graphene, Technical University of Denmark, DK-2800 Kgs. Lyngby, Denmark}

\author{N. Asger  Mortensen}
\affiliation{Department of Photonics Engineering, Technical University of Denmark, DK-2800 Kgs. Lyngby, Denmark}
\affiliation{Center for Nanostructured Graphene, Technical University of Denmark, DK-2800 Kgs. Lyngby, Denmark}
\affiliation{Institute of Applied Physics, Abbe Center of Photonics, Friedrich Schiller University Jena, Germany}

\date{\today}
%

\begin{abstract}\color{black!88!white}
Quantum effects of plasmonic phenomena have been explored through ab-initio studies, but only for exceedingly small metallic nanostructures, leaving most experimentally relevant structures too large to handle. We propose instead an effective description with the computationally appealing features of classical electrodynamics, while quantum properties are described accurately through an infinitely thin layer of dipoles oriented normally to the metal surface. The nonlocal polarizability of the dipole layer is mapped from the free-electron distribution near the metal surface as obtained with 1D quantum calculations, such as time-dependent density-functional theory (TDDFT), and is determined once and for all. The model can be applied to any system size that is tractable within classical electrodynamics, while capturing quantum plasmonic aspects of nonlocal response and a finite work function with TDDFT-level accuracy. Applying the theory to dimers we find quantum-corrections to the hybridization even in mesoscopic dimers as long as the gap is sub-nanometric itself.
\end{abstract}

\maketitle

\color{black!88!white}

Nanoscale metallic structures receive considerable attention due to plasmon phenomena beyond classical electrodynamics~\cite{Scholl:2012,Raza:2012b,Savage:2012,Tan:2014,Ciraci:2012b}. In deep nanoscale structures, quantum effects will manifest themselves~\cite{Tame:2013}. Size-dependent plasmon energies of nanoscale metallic particles, an accustomed topic in cluster science and surface physics~\cite{Feibelman:1982,Liebsch:1993,Apell:1983,Tigge:1993,Charle:1998}, has been revisited with new insights \cite{Scholl:2012,Raza:2012b}.
Dimers hosting \AA ngstrom-scale gaps are being investigated intensively in search of quantum-tunneling effects~\cite{Savage:2012,Tan:2014}, though interpretations highlight both quantum tunneling~\cite{Esteban:2012} and semi-classical dynamics~\cite{Mortensen:2014}. Time-dependent density-functional theory (TDDFT) allows explorations into the quantum regime~\cite{Nordlander:2012,Teperik:2013,Stella:2013}, albeit limited to atomic-scale systems.

Here we focus instead on the mesoscopic scale that is typical for state-of-the-art experiments: too small to obey the classical local-response approximation (LRA), but too large for an efficient quantum treatment. Plasmonics at this scale has so far been described with some success by semi-classical hydrodynamic Drude models (HDM) that exhibit nonlocal response~\cite{Ciraci:2012b,Raza:2011,Mortensen:2014,Luo:2013,David:2014,Raza:2015b} and by the quantum-corrected model (QCM) for gaps between two metals~\cite{Esteban:2012}. It is however desirable to look beyond these models.

\begin{figure}
\centering
\includegraphics[width=7cm]{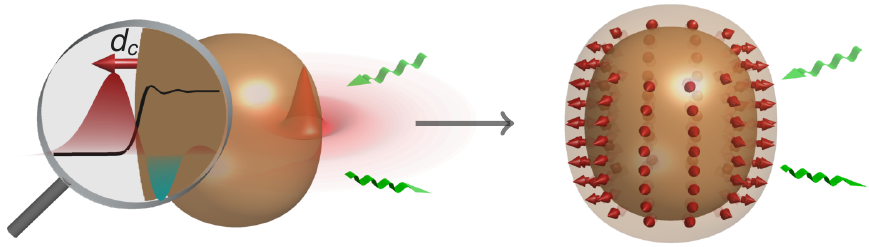}
\caption{In the projected-dipole model (PDM), light-driven quantum-plasmonic response in a metallic nanostructure (left) is represented in classical local-response electrodynamics by an infinitely thin layer of dipoles that point perpendicularly to the surface (right). The loupe (left) highlights the spill-out of the microscopic electron distribution inducing a dipole
moment proportional to the Feibelman parameter $d_c$.}
\label{fig:1}
\end{figure}

In this Letter, we present the projected-dipole model (PDM), a theory that describes optical properties of mesoscopic plasmonic systems with TDDFT level accuracy. Their quantum effects are captured by a zero-thickness projected-dipole layer (PDL) onto the surface of the metal, see Fig.~\ref{fig:1}. The idea is partly inspired by Ref.~\onlinecite{Luo:2013}, where a {\em finite}-thickness local layer represents nonlocal response within the HDM. Here, we go beyond the semi-classical HDM and reproduce quantum phenomena of a microscopic theory of choice, TDDFT. A crucial computational advantage of the PDL is that its inclusion can be elegantly absorbed into boundary conditions at the interface.

\emph{Projected-dipole layer.---} We first introduce the PDL, the key theoretical concept of our approach. To reproduce a quantum calculation of the  reflectivity of a planar metal-air interface at $x=0$ within a classical model, we add an infinitely thin layer of dipoles to the corresponding classical interface.  The electric response of this PDL is
$\mathbf P_{\rm PDL}=\epsilon_0 \alpha\hat {\mathbf n}\hat {\mathbf n}\cdot\mathbf E_{\rm air}(x=0)$, 
where $\hat {\mathbf n}$ is the normal vector pointing outward the metal surface and  $\mathbf E_{\rm air}$ is the electric field in the air region. The polarization field $\mathbf P_{\rm PDL}$ is a projection of the electric dipole onto the surface normal, with a yet unknown polarizability $\alpha$ that we identify uniquely below, after further motivating  our model.
The semi-infinite metal at $x<0$ that is in the electrostatic limit driven by an external electric potential $\phi_{\rm ext}={\mathrm e}^{{\mathrm i}{\mathbf k_{\scriptscriptstyle  ||}\cdot \mathbf s}+k_{\sbp}x}$, where $\mathbf s=y\hat{\mathbf y}+z \hat {\mathbf z}$,
exhibits an induced electron density $\rho(x)$ that within the LRA is a Dirac delta function $q_1\delta(x)$. The induced electric potential outside the metal equals $-e q_1{\mathrm e}^{-k_{\sbp}x}/2k_{\sbp}$.
By contrast, $\rho(x)$ spreads across the geometrical boundary in quantum theories. The corresponding induced electric potential outside the metal consequently has a multipole expansion that starts as $(-eq_2{\mathrm e}^{-k_{\sbp}x}/2k_{\sbp})[1+k_{\sbp}d_c+\mathcal{O}(k_{\sbp}^2d_c^2)]$. Here the first and second terms in the bracket
represent the monopole and dipole contributions, respectively, with $q_2=\int dx\, \rho(x)$ the total induced electron charge and $d_c=\int dx\, x \rho(x)/q_2$ the Feibelman parameter, i.e., the centroid of the induced charge.
Clearly, the leading-order quantum effect is to induce a dipole moment proportional to $d_c$ in the surface region as indicated in Fig. \ref{fig:1}. This motivates our choice of a PDL within classical electrodynamics to model quantum effects.

\emph{Planar-surface polarizability.---} Next we outline in more detail how to identify  the polarizability $\alpha$ of the PDL. Within TDDFT, we solve $\rho(x,\omega)$ self-consistently from
\begin{align}
\rho(x,\omega){=}\!{\int\!\!dx'\,\left[\Xi (x,x',\omega)\rho (x,\omega){-}e{{\,{\chi _{\scriptscriptstyle\rm KS}}(x,x',\omega )\phi_{\rm ext} (x',\omega)}}\right]}.\nonumber
\end{align}
Here, $\chi _{\scriptscriptstyle\rm KS}$ represents the Kohn--Sham response function and $\Xi$ accounts for Coulomb and exchange-correlation interactions \cite{Gross:2006}. For simplicity we employ the jellium approximation of the positive-ion background and the local-density approximation of the exchange-correlation potential.
 
We calculate the TDDFT reflection amplitude
$R_{\scriptscriptstyle\rm KS}(k_{\sbp},\omega)=-\frac{e}{2k_{\sbp}}\int\!dx\,\rho(x,\omega){\mathrm e}^{k_{\sbp}x}$, and require that the classical PDL problem gives the same value, which gives
\begin{align}
\alpha(\omega,k_{\sbp})=\frac{[1-R_{\scriptscriptstyle\rm KS}(\omega,k_{\sbp})]-\epsilon_{\rm m}^{\rm LRA}[1+R_{\scriptscriptstyle\rm KS}(\omega,k_{\sbp})]}{k_{\sbp}\epsilon_{\rm m}^{\rm LRA}(\omega)[1-R_{\scriptscriptstyle\rm KS}(\omega,k_{\sbp})]},
\label{eq:alphaNL}
\end{align}
where $\epsilon_{\rm m}^{\rm LRA}$ is the bulk permittivity of the metal. To interpret this result, we first consider its long-wavelength (or $k_{\sbp} \to 0$) limit $\alpha_{\rm L}$. Since $R_{\scriptscriptstyle\rm KS}$ in this limit equals $-[\epsilon_{\rm m}^{\rm LRA}(\omega)-1][1+k_{\sbp}d_c(\omega)]/[\epsilon_{\rm m}^{\rm LRA}(\omega)+1-(\epsilon_{\rm m}^{\rm LRA}(\omega)-1)k_{\sbp}d_c(\omega)]$ \cite{Liebsch:1991},  we obtain
$\alpha_{\rm L}=
[\epsilon_{\rm m}^{\rm LRA}(\omega)-1] d_c(\omega)/\epsilon_{\rm m}^{\rm LRA}(\omega)$,
which indeed shows a direct relation of the PDL polarizability with
$d_c$ as our intuitive expectation.
The latter parameterizes two important quantum effects: electron spill-out via ${\rm Re} [d_c]$ and electron-hole excitations via ${\rm Im} [d_c]$~\cite{Feibelman:1982}. Within the HDM, $d_c$ becomes $-1/k_{\rm L}$ (``spill-in'', proper for noble metals \cite{Raza:2012b,Ciraci:2012b} but not for simple metals \cite{Teperik:2013,Stella:2013}) in terms of the Thomas--Fermi
wavenumber $k_{\rm L}$~\cite{Raza:2011}, and $\alpha_{\rm L}$ becomes $(\epsilon_{\rm m}^{\rm LRA}-1)/(\epsilon_{\rm m}^{\rm LRA}k_{\rm L})$, which has the same effect as the local dielectric layer in Ref.~\onlinecite{Luo:2013}. The standard HDM cannot describe electron spill-out or electron-hole excitation, though; only recent HDM extensions do include spill-out~\cite{David:2014,Toscano:2014,Yan:2015}.

\begin{figure}
\centering
\includegraphics[width=7.5cm]{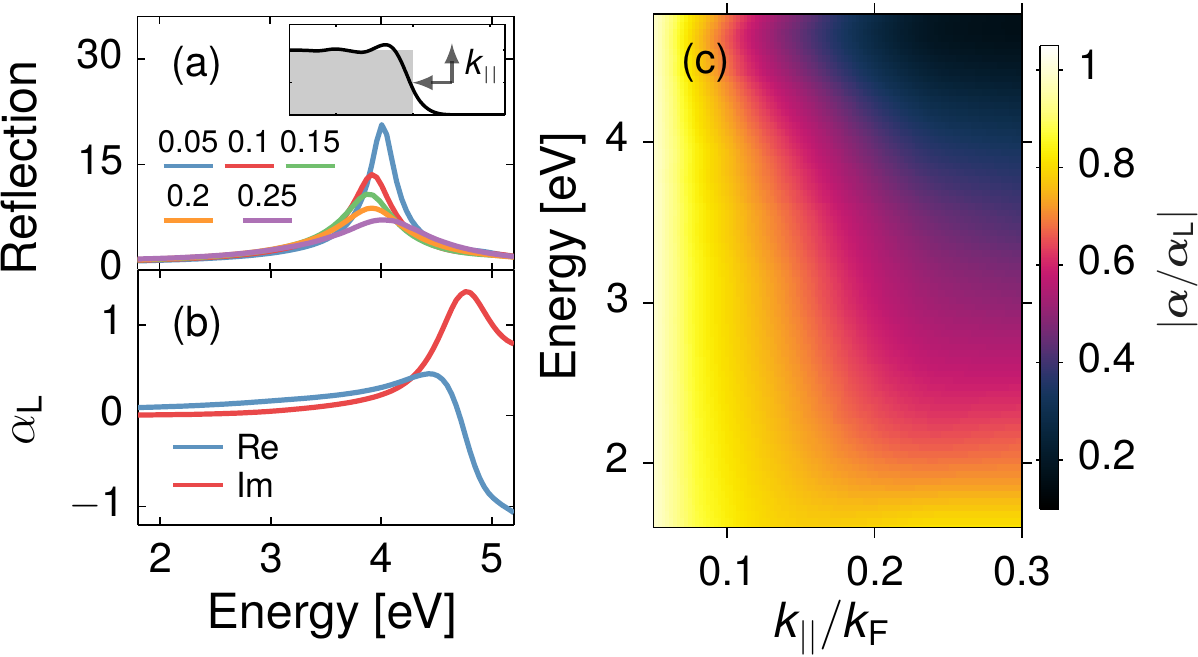}
\caption{(a) TDDFT reflection spectra of a Sodium surface for wave incidence with different $k_{\scriptscriptstyle ||}$ (in units of $k_{\rm F}$). (b) Long-wavelength polarizability $\alpha_{\rm L}$. (c) Magnitude of polarizability $\alpha$ normalized by $\alpha_{\rm L}$.
}
\label{fig:2}
\end{figure}

The plasmonic response of Sodium exhibits an interplay between nonlocal response and quantum spill-out due to a finite work function, and illustrates the potential of our PDM.
Fig.~\ref{fig:2}(a) shows reflection spectra off a planar Sodium-air interface for different $k_{\sbp}$, a standard TDDFT calculation. The peak around $4$\,eV corresponds to the surface plasmon (SP) resonance, which first shifts to the red and then to the blue as $k_{\sbp}$ increases, while the width shows a noticeable broadening due to enhanced electron-hole excitations. Fig.~\ref{fig:2}(b) depicts $\alpha_{\rm L}$, with around $4.8$\,eV  a sign change of ${\rm Re} [\alpha_{\rm L}]$ and a corresponding peak in ${\rm Im} [\alpha_{\rm L}]$. This multipole surface plasmon (MSP) resonance~\cite{Liebsch:1993}, with dipole-like charge distribution at the surface,  is barely visible in Fig.~\ref{fig:2}(a) due to the nearby stronger SP (monopole) resonance. Fig.~\ref{fig:2}(c) shows an increased $k_{\sbp}$-dependence of $\alpha$ for larger energies, so $\alpha_{\rm L}$ is a good approximation of $\alpha$ only at low energies.

\emph{Arbitrary structures.---} With the flat-layer polarizability $\alpha(\omega,k_{\sbp})$ defined in Eq.~(\ref{eq:alphaNL})--the input parameters of the theory, we here generalize it to an arbitrary metal structure with a locally curved surface.
Thus, we seek the polarizability in real space, which for the planar surface is given by the inverse Fourier transform of $\alpha(\omega,k_{\sbp})$, which assumes the nonlocal form $\alpha_{\rm p}(\omega,|\mathbf s-\mathbf s'|)$ that only depends on the distance between the points on the surface. As observed in Fig.~\ref{fig:2}(c), $\alpha$ is nearly constant for $k_{\sbp}<k_{c}$ with $k_c\approx 0.1k_F$. Thus the nonlocal range of the flat-layer $\alpha_{\rm p}$ is of the order $1/k_c\approx 1\,\rm nm$. For a general curved surface as long as the local curvature does not exceed $1$ nm$^{-1}$, we make the approximation that the polarizability of a curved surface has the expression of a planar surface, with the distance on the curved surface defined as the length of the geodesic~\cite{Ulf:2009}. We thus obtain
\begin{align}
\alpha(\omega,\mathbf s, \mathbf s') &=\sum\limits_g \alpha_{\rm p}(\omega,|\mathbf s-\mathbf s'|_g).
\label{eq:alphaNLRS}
\end{align}
The sum over $g$ runs over all geodesics between $\mathbf s$ and $\mathbf s'$, with length  $|\mathbf s-\mathbf s'|_g$. When $g=1$ and the geodesic is a straight line, Eq.~(\ref{eq:alphaNLRS}) simplifies to the planar-surface case.

By using  Eq.~(\ref{eq:alphaNLRS}) for $\alpha$, the response of an arbitrary structure including quantum effects can be predicted within classical electrodynamics. With retardation included, the presence of the PDL is equivalent to the boundary condition
\begin{align}
\mathbf E_{{\scriptscriptstyle ||},\,\rm air}(\mathbf s)-\mathbf E_{{\scriptscriptstyle ||},\,\rm metal}(\mathbf s)=-\boldsymbol\nabla_{\mathbf s}\int d\mathbf s' \alpha(\omega,\mathbf s,\mathbf s')\hat {\mathbf n}(\mathbf s')\cdot\mathbf E_{\rm air}(\mathbf s'),
\label{eq:PDMBC}
\end{align}
where the subscript "$||$" represents the parallel components of the electric fields~\cite{SeeSupplementalMaterial}. This boundary condition can be conveniently adopted in numerical computations. If large-$k_{\sbp}$ fields do not contribute to the dynamics significantly or if one is just interested in the leading-order quantum effects, $\alpha$ can be approximated by the local form
$\alpha_{\rm L}(\omega)\delta(\mathbf s-\mathbf s')$,
as can be derived by replacing $\alpha(k_{\sbp},\omega)$ with $\alpha_{\rm L}(\omega)$. The boundary condition simplifies to $\mathbf E_{{\scriptscriptstyle ||},\,\rm air}-\mathbf E_{{\scriptscriptstyle ||},\,\rm metal}=-\alpha_{\rm L}\boldsymbol\nabla\left(\hat {\mathbf n}\cdot\mathbf E_{\rm air}\right)$. Further, beyond the local approximation, with the observation that the nonlocal polarizability is a short-range function, we can make the shortest-geodesic approximation, i.e., just keeping the term of the shortest geodesic length in Eq.~(\ref{eq:alphaNLRS})~\cite{SeeSupplementalMaterial}.

In the following we focus on plasmonic structures with 1D translational invariance, in the $z$-direction say, which include extended 1D gratings and arbitrarily shaped nanowires. This simplifies the task of finding geodesics.
After Fourier transformation in the $z$-direction with wavenumber $k_z$,  the polarizability of the PDL in Eq.~(\ref{eq:alphaNLRS}) becomes
\begin{align}
\alpha(\omega,k_{z}, l, l') = \sum\limits_{n =  - \infty }^{ + \infty } {\alpha(k_n,\omega)\mathrm e^{\scalebox{0.9}{$\mathrm i\frac{2\pi n}{\rm L}|l-l'|$}}}.\nonumber
\end{align}
Here $\rm L$ is the total arc length of the boundary in the $x,y$-plane, $l\in[0\, ,\,\rm L]$ is the local coordinate of the boundary with Cartesian metric, and $k_{n}=[(2\pi n/L)^2+k_z^2]^{1/2}$. In this picture, fields propagate along the curved boundary as if it were a planar interface with artificial periodicity $\rm L$, so that $\alpha$ is spanned by the associated Fourier components.

\begin{figure}
\centering
\includegraphics[width=7.5cm]{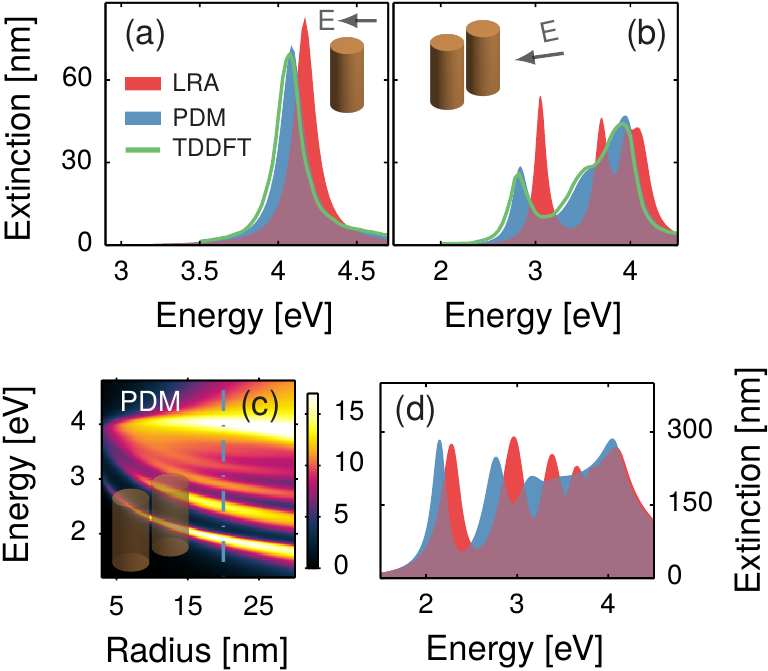}
\caption{Extinction properties of Sodium nanowires within LRA, PDM, and TDDFT. (a) Cylindrical wire with radius $4.9\,\rm nm$ and (b) corresponding dimer with a gap of $0.74\,\rm nm$. (c) Dimer structure with radius varying from 3 nm to 30 nm and 0.74 nm gap distance, where the extinction cross sections are normalized by the nanowire radius, and (d) 20 nm radius case [dash-dotted line in (c)].
TDDFT results in (a,b) are reproduced from Ref.~\onlinecite{Teperik:2013}.}
\label{fig:3}.
\end{figure}

\emph{Benchmarks against TDDFT.---} We will confirm the validity of the PDM for a Sodium cylinder of radius $4.9\,\rm nm$ (i.e. curvature assumption behind PDM is fulfilled), excited by a plane wave incident normally to the wire axis and with an in-plane polarized electric field.
The local bulk permittivity is described by the Drude model, $\epsilon_{\rm metal}^{\rm LRA}=1-\omega_{\rm p}^2/\omega(\omega+i\gamma)$, where $\omega_{\rm p}=5.89\,\rm eV$. For the damping we use $\gamma=0.16\,\rm eV,$ and 0.085\,eV for the LRA and PDM, respectively, to match the resonance width.
The extinction spectra predicted by LRA, PDM, and TDDFT are presented in Fig.~\ref{fig:3}(a). The LRA and PDM spectra are calculated with the Green's function surface-integral method~\cite{Yan:2013}, the TDDFT results are reproduced from Ref.~\onlinecite{Teperik:2013}. Since retardation is neglected in the TDDFT, the LRA and PDM are also calculated in the electrostatic limit. The PDM and TDDFT spectra in Fig.~\ref{fig:3}(a) agree extremely well, and both predict that the SP peak around $4\,\rm eV$ is redshifted with respect to the LRA peak.
Next, we consider a dimer of such cylindrical nanowires, separated by a gap of only $0.74$ nm. Fig.~\ref{fig:3}(b) shows a pronounced discrepancy between the LRA extinction spectrum on the one hand and the PDM and TDDFT spectra on the other, with nonclassical broadening in the latter two. Moreover, as one of our main results we find that the PDM and TDDFT spectra agree well, quite surprisingly given the fact that the PDM models both metal-air interfaces in the gap region in an independent-surface approximation (ISA).

Quantum effects are known to be negligible for large-scale structures, e.g. nanowires above 10 nm radius. The multi-scale dimer system with a sub-nanometer gap distance introduces an additional scale for confining plasmon fields and enhances the quantum effects~\cite{Nordlander:2012,Raza:2015}. Such a setup is usually far too challenging for ab-initio studies. However, the PDM can be applied to explore the quantum effects of such a system. Fig.~\ref{fig:3}(c) depicts the color map of the extinction cross section (normalized by wire radius) predicted by the PDM for the dimer structure as in Fig.~\ref{fig:3}(b) with radius varying from 3 nm to 30 nm. Retardation effects are now also included. To demonstrate the quantum effects of a typical multi-scale system, Fig.~\ref{fig:3}(d) plots the extinction spectra at 20 nm radius of Fig.~\ref{fig:3}(c) (dash-dot line), contrasting PDM and LRA results. Similarly as in Fig.~\ref{fig:3}(b), quantum effects manifest themselves by shifting and broadening the SP resonances. Interestingly, the higher-order modes in this case are less affected by quantum effects, e.g., the quadrupole mode (around 2.8 eV) and hexapole mode (around 3.2 eV) show clear peaks in the PDM spectra, since plasmon fields in larger structures are less confined.

\emph{More Gap Effects.---} In the PDM, we invoked the ISA and thus only included quantum effects related to the single metal surface.
When the gap approaches the regime set by the $d_c$ length scale,
it also implies that a significant wavefunction overlap of the two surfaces should be taken into account.
An intuitive way to do this is to treat the gap as an effective medium, as firstly introduced in the QCM with a local Drude permittivity constructed to mimic the dissipation associated with quantum tunneling (QT)~\cite{Esteban:2012}. To improve our PDM method for the smallest gaps, here we present an unambiguous extraction of the effective gap permittivity directly from TDDFT calculations, instead of relying on the assumption that QT is the cause of all gap-related dissipation.

\begin{figure}
\centering
\includegraphics[width=8.2cm]{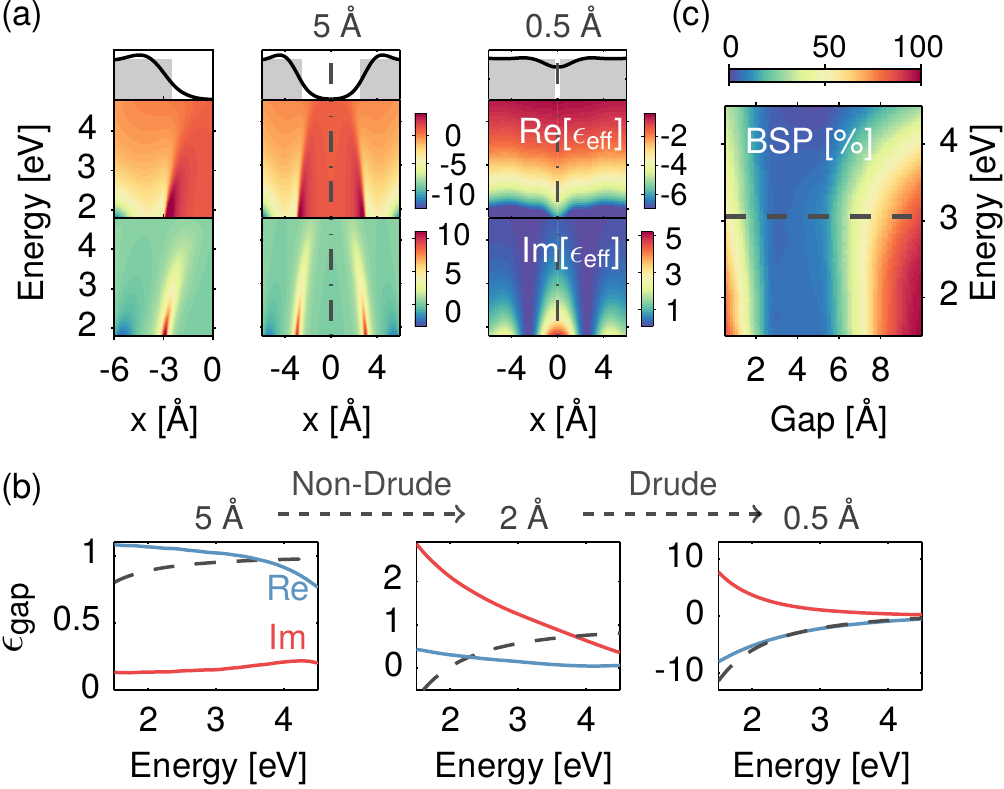}
\caption{Gap effects for planar Sodium-vacuum-Sodium gap structures within TDDFT.   (a) Effective permittivity for single interface and for 5{\AA} and 0.5{\AA} gaps. (b) Frequency dependence of the corresponding mid-gap effective permittivities. (c) Electron-hole backward scattering proportion (BSP) as a function of gap distance and energy.}
\label{fig:4}
\end{figure}

Turning from the spectral to the spatial information provided by TDDFT calculations, we extract the effective local gap permittivity $\epsilon_{\rm eff}$, defined as $\rm D/[\epsilon_{0}\rm E]$. In Fig.~\ref{fig:4}(a) we visualize $\epsilon_{\rm eff}$ for a single interface and for two gap distances, of 5{\AA} and 0.5{\AA}. For the 5{\AA} gap case, two separated metallic systems can be identified as seen from the clear boundaries between the negative (metal) and positive (gap) valued regions of ${\rm Re}[\epsilon_{\rm eff}]$, as for the single interface. Furthermore, the loss ${\rm Im}[\epsilon_{\rm eff}]$ clearly peaks right at the interfaces where we likewise have ${\rm Re}[\epsilon_{\rm eff}]\sim 0$. In other words, we find that dissipation (associated with electron-hole excitations) occurs around the metal surface, in agreement with mesoscopic transport theory~\cite{Datta:1995}, rather than in the vacuum region itself. By contrast, for the 0.5{\AA} gap no definite boundary of the two metals can be observed from the equilibrium electron density distribution. The induced electron current at the gap center is accordingly the classical Drude current. In the optical spectra this regime (gaps of 1--2{\AA} and below) is characterized by a transition from the bonding-dipole plasmon (BDP) resonance to the charge-transfer plasmon (CTP). 

Fig.~\ref{fig:4}(b) shows mid-gap properties, $\epsilon_{\rm gap}=\epsilon_{\rm eff}(x=0)$, for gaps of 5{\AA}, 2{\AA}, and 0.5{\AA}. In general, we anticipate a transition from non-Drude (quantum) response to Drude (classical) response as the gap reduces. Indeed, for the 0.5{\AA} gap the mid-gap response is well captured by the classical Drude model (dashed line) using a plasma frequency determined by the mid-gap equilibrium electron density~\cite{Ichikawa:2011}. On the other hand, the larger separation of 5{\AA} shows a complete non-Drude behavior. The 2{\AA} gap exhibits almost frequency independent ${\rm Re}[\epsilon_{\rm gap}]$, thus constituting the cross-over gap size between non-Drude and Drude responses.

\begin{figure}
\centering
\includegraphics[width=7.5cm]{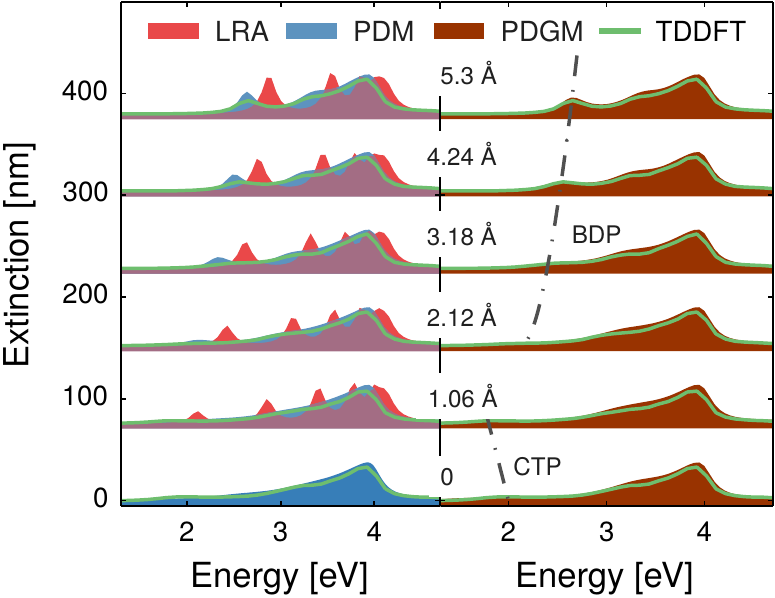}
\caption{Extinction cross sections of a cylindrical Sodium nanowire dimer with radius $4.9\,\rm nm$, for gaps decreasing from $5.3$ to $0\AA$. Comparison of predictions by LRA, TDDFT, PDM, and PDGM. TDDFT spectra originate from Ref.~\onlinecite{Teperik:2013}.}
\label{fig:5}
\end{figure}

\emph{Modeling {\AA}-sized gaps.---}  We will now discuss how in the PDM a dimer gap can be treated as an effective medium
(in spirit close to the QCM~\cite{Esteban:2012}), thereby accounting for not only  quantum effects of both individual metal interfaces (as before, employing the ISA) but also for their strong mutual proximity. The procedure in this Projected-Dipole Gap Model (PDGM) is to extract $\epsilon_{\rm gap}$ as before from TDDFT calculations of a planar dimer, and to treat the gap as an effective medium with permittivity $\epsilon_{\rm gap}$ within the PDM framework. To illustrate the accuracy of the PDGM,
we consider a nanowire dimer of two Sodium cylinders with radius 4.9\,nm driven by a uniform electric field along the gap. In Fig.~\ref{fig:5} we illustrate the extinction spectra according to LRA, PDM, PDGM, and TDDFT, as the gap distance decreases from 5.3 to 0{\AA} (contact). 
As expected, not all LRA resonances are supported by the more accurate TDDFT calculations. The PDM captures the resonance broadening as the gap is reduced, with spectra closely resembling those of TDDFT. However, the PDM neglects strong gap effects, and therefore underestimates the resonance broadening. For the same reason the CTP does not appear before physical contact. Turning to the PDGM, we find almost perfect agreement with the TDDFT spectra (both in terms of resonance positions and broadening) even for the smallest gaps. 

\emph{Discussion.---}
The great accuracy of the PDM in Fig.~\ref{fig:3} and of the PDGM in Fig.~\ref{fig:5} calls for a discussion of the dissipative quantum processes at interfaces and near gaps. The TDDFT calculations account for plasmon damping due to electron-hole (e-h) excitations.
To interpret our results, we distinguish two scattering processes (details in Supplementary Material~\cite{SeeSupplementalMaterial}): an electron state that propagates towards the interface or gap can create e-h pairs in forward- or in backward scattering. For a single interface, at energies below the work function all dissipation due to e-h creation is associated with backward scattering, while forward scattering can only exist  above the work function. By contrast for a dimer structure with a narrow gap, forward scattering below the work function may also play a role. The PDM assumes all gap- and interface-related loss to be due to backward scattering, whereas the QCM calculates all gap-related loss as due to QT, i.e. forward scattering~\cite{Esteban:2012}. Finally, the PDGM does not systematically exclude one of both processes.
So what dominates e-h creation, forward or backward scattering?

To make our discussion specific, we consider a vacuum gap formed between two parallel planar Sodium surfaces driven by a uniform electric field. Fig.~\ref{fig:4}(c) depicts the e-h backward scattering proportion (BSP)~\cite{SeeSupplementalMaterial}, defined by the ratio between the backward and total scattering rates. The BSP exhibits completely different behavior for gap sizes above and below 3{\AA}. For gaps $> 7{\AA}$, the BSP approaches $100\%$ below the work function, which explains the validity of the PDM in Fig.~\ref{fig:3}. For smaller gaps down to 3{\AA}, the BSP continuously decreases. Below the work function (3.06 eV for Sodium interface) this is due to QT. The latter mechanism also explains the abrupt decrease of the BSP as the energy exceeds the work function [dashed line in Fig.~\ref{fig:4}(a)]. As the gap shrinks further below 3{\AA}, the concepts of QT and work function are not well defined, since the vacuum gap separating different systems cannot be unambiguously defined. Thus, the traditional tunneling picture is not proper. This is evident from the observation that in Fig.~\ref{fig:4}(c) the BSP starts {\em increasing} when making gaps smaller than 3{\AA}, and eventually dominates the forward-scattering at low energies.


In conclusion, we propose a projected-dipole model (PDM) for quantum plasmonics, where the complex dynamics of the quantum electron gas is explored with TDDFT (or other ab-initio methods) in the vicinity of the surface and then subsequently the quantum response is represented within the common framework of classical local-response electrodynamics as a sheet of dipoles pointing normally to the surface. Remarkably, we have shown that this mapping captures quantum plasmonic aspects of nonlocal response and a finite work function with TDDFT-level accuracy. Our work also provides key insight into the possible importance of optical tunneling currents in plasmonic dimers. Finally, our results have great potential for multi-scale modelling in computational plasmonics, where quantum effects can now be accurately explored  even in large systems, while enjoying the efficiency and accuracy of state-of-the-art electromagnetic solvers for classical local-response electrodynamics.

The Center for Nanostructured Graphene is sponsored by the Danish National Research Foundation, Project DNRF58. The work was also supported by the Danish Council for Independent Research--Natural Sciences, Project 1323-00087, and by the Lundbeck Foundation, grant no. 70802. N.~A.~M. acknowledges the Proexcellence program of the Thuringian State Government (ACP2020).


%

\newpage
\newpage

\end{document}